\documentclass[journal]{IEEEtran}
\usepackage[T1]{fontenc}
\usepackage{float}
\usepackage{amsmath}
\usepackage{amsthm}
\usepackage{amssymb}
\usepackage{stackrel}
\usepackage{graphicx}
\usepackage{setspace}
\usepackage{url}
\allowdisplaybreaks

\usepackage{caption}
\usepackage{subfigure}
\usepackage{algorithm}
\usepackage[graphicx]{realboxes}
\usepackage{algorithmic}
\usepackage{verbatim}

\makeatletter

\floatstyle{ruled}
\newfloat{algorithm}{tbp}{loa}
\providecommand{\algorithmname}{Algorithm}
\floatname{algorithm}{\protect\algorithmname}
\theoremstyle{plain}

\theoremstyle{definition}

\theoremstyle{plain}

\theoremstyle{plain}
\newcommand{\RNum}[1]{\uppercase\expandafter{\romannumeral #1\relax}}
\usepackage[thicklines]{cancel}
\IEEEoverridecommandlockouts
\usepackage{ifpdf}
\usepackage{cite}
\usepackage{graphicx}
\ifCLASSOPTIONcompsoc
\else
\fi
\usepackage{graphicx}
\usepackage{epstopdf}
\usepackage{mathtools}
\usepackage{booktabs} 
\usepackage{multirow}
\usepackage{setspace}
\usepackage{lettrine} 
\usepackage{bm}
\makeatother
\usepackage{babel}

\usepackage{amsmath}

\usepackage{hyperref}
\hypersetup{
 colorlinks=true,
 linkcolor=black,
 citecolor=black,
 filecolor=black,
 urlcolor=black,
 }

\usepackage{booktabs}
\usepackage{longtable}
\usepackage{amsthm}
\usepackage{enumitem,textcomp,amsfonts,multicol,epsfig,bbding,cases,dsfont,bbm,tabularx,makecell,lscape,chngcntr,mwe,adjustbox}
\usepackage{array}
\usepackage{mdwmath}
\usepackage{mdwtab}
\usepackage{eqparbox}
\usepackage{upgreek}
\usepackage{stfloats}
\usepackage{afterpage}
\usepackage{mathrsfs}

\begin{document}
\captionsetup[figure]{font={small}, name={Fig.}, labelsep=period}
\title{A Learnable SIM Paradigm: Fundamentals, Training Techniques, and Applications}
    \author{
    Hetong Wang, Yashuai Cao, and~Tiejun Lv,~\IEEEmembership{Senior Member,~IEEE}

    \thanks{Manuscript received 3 November 2025; revised 25 February 2026, and 6 March 2026; accepted 12 March 2026. This paper was supported in part by the National Natural Science Foundation of China under No. 62271068.
    (\emph{corresponding author: Tiejun Lv}.)}

    \thanks{
    H. Wang and T. Lv are with the School of Information and Communication Engineering, Beijing University of Posts and Telecommunications (BUPT), Beijing 100876, China (e-mail: \{htwang\_61, lvtiejun\}@bupt.edu.cn)
    }
    \thanks{
    Y. Cao is with the School of Intelligence Science and Technology, University of Science and Technology Beijing, Beijing 100083, China (e-mail: caoys@ustb.edu.cn).
    }
}

\maketitle
\begin{abstract}
Stacked intelligent metasurfaces (SIMs) represent a breakthrough in wireless hardware by comprising multilayer, programmable metasurfaces capable of analog computing in the electromagnetic (EM) wave domain.
By examining their architectural analogies, this article reveals a deeper connection between SIMs and artificial neural networks (ANNs).
Leveraging this profound structural similarity, this work introduces a learnable SIM architecture and proposes a learnable SIM-based machine learning (ML) paradigm for sixth-generation (6G)-and-beyond systems.
Then, we develop two SIM-empowered wireless signal processing schemes to effectively achieve multi-user signal separation and distinguish communication signals from jamming signals.
The use cases highlight that the proposed SIM-enabled signal processing system can significantly enhance spectrum utilization efficiency and anti-jamming capability in a lightweight manner and pave the way for ultra-efficient and intelligent wireless infrastructures.
\end{abstract}


\section{Introduction}\label{Sec:I}
\IEEEPARstart{T}{he} advent of stacked intelligent metasurfaces (SIMs) marks a transformative leap beyond conventional reconfigurable intelligent surfaces (RISs).
This three-dimensional metamaterial, composed of multiple transmissive programmable metasurfaces~\cite{10515204,10279173}, greatly increases available degrees of freedom (DoFs) and enables analog computing directly within the electromagnetic (EM) wave domain.
Consequently, a SIM can serve as both a transmit precoder and a receiver combiner~\cite{10534211}, processing multiple data streams in parallel at the speed of light. \par

\subsection{The Role of SIMs in 6G-and-Beyond}
As a wave-based analog computing platform, the SIM dramatically improves processing efficiency and paves the way for transceiver designs with simplified hardware~\cite{10515204}, leading to substantial reductions in computational cost, hardware complexity, processing latency and power consumption.
\begin{itemize}
    \item \textbf{Reduced computational cost}: The SIM physically approximates end-to-end quasi-orthogonal multiple-input multiple-output (MIMO) channels directly in the EM wave domain.
    This delivers pre-separated signal streams to the receiver, avoiding the need for complex digital baseband operations such as matrix inversion and decomposition.
    \item \textbf{Simplified hardware architecture}: By directly providing pre-separated signal streams, the SIM allows the baseband to operate with low-precision, low-power digital-to-analog (DAC)/analog-to-digital (ADC) converters~\cite{10515204} thanks to enhanced quantization noise tolerance. The large-scale array gain of the SIM further reduces the number of required radio frequency (RF) chains to match the user data streams.
    \item \textbf{Lower processing latency}: Multiple data streams are processed in parallel at the speed of light within the EM wave domain, bypassing the sequential delays inherent to digital signal processing in traditional MIMO systems.
    \item \textbf{Improved energy efficiency (EE)}: The SIM's simplified hardware architecture substantially reduces digital baseband power consumption. Through wave-domain multi-user interference (MUI) management, the SIM replaces the hardware-intensive, power-consuming components in conventional multi-user multiple-input single-output (MU-MISO) systems.
\end{itemize} \par

These inherent advantages establish SIMs as a pivotal technology for sixth-generation (6G)-and-beyond, addressing challenges like high-frequency path loss, the energy footprint of dense networks, and growing computational demands~\cite{cao2024intelligent}. Thus, SIMs enable a shift toward realizing future wireless systems that are ubiquitous, ultra-reliable, low-latency, and massively connected, in an energy- and cost-efficient manner~\cite{10596064}. \par
\begin{figure*}[t]
	\centering{}\includegraphics[width=0.95\textwidth]{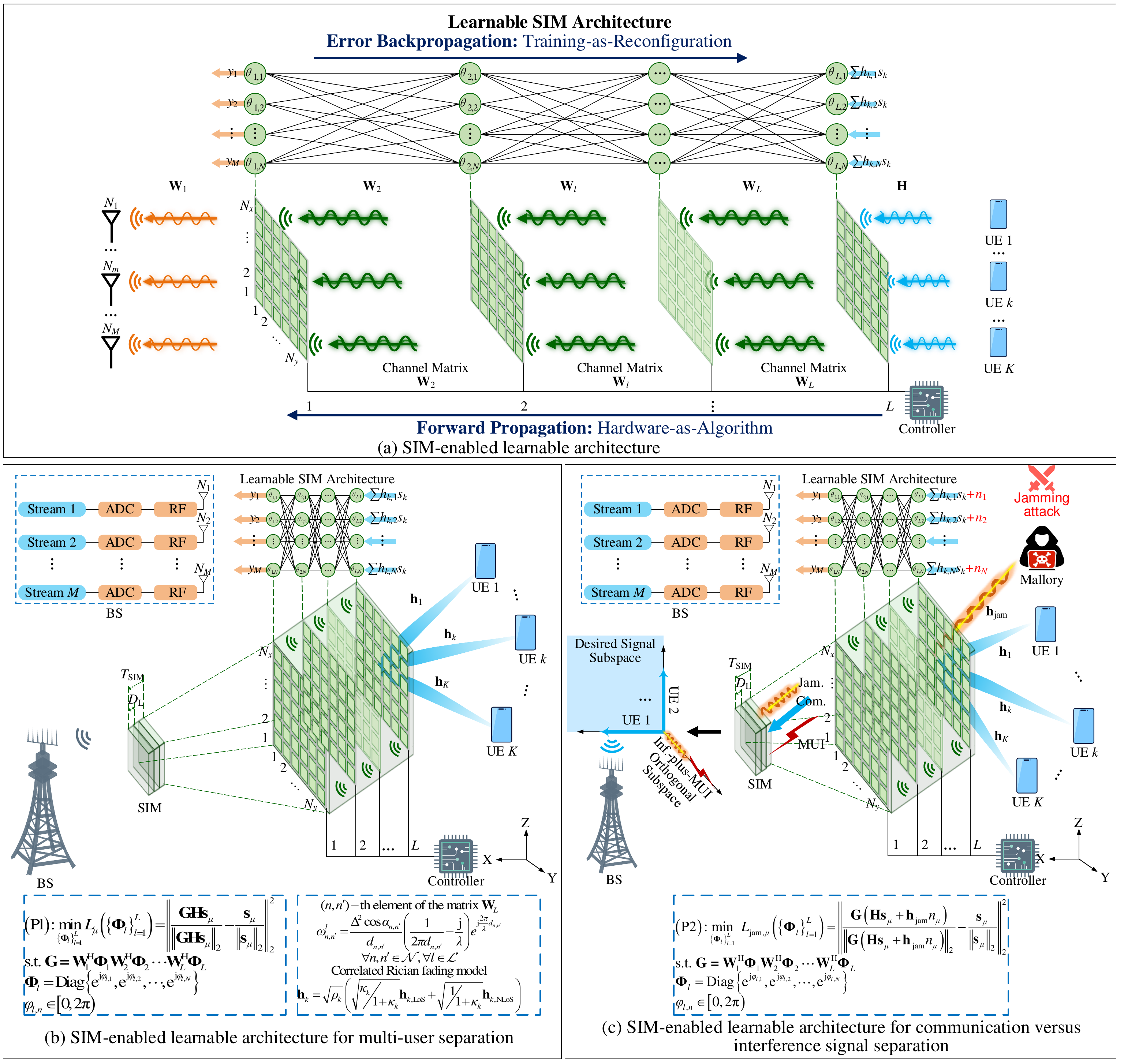}
	\caption{ML paradigm with the SIM-enabled learnable architecture including: (a) The SIM provides a physical platform for implementing a learnable architecture inspired by ANNs; (b) The SIM is integrated with the BS as the combiner to effectively distinguish multi-user transmission signals in MU-MISO uplink systems; (c) The learnable SIM can be further designed to separate communication and jamming signals, and defend against Mallory's jamming attack.}
	\label{fig:II.A.1}
\end{figure*}

\subsection{Contribution and Organization}
The innate analog computing capacity of SIMs enables diverse advanced applications, such as managing MUI in MU-MISO systems as integrated precoders/combiners~\cite{10596064}, compensating power amplifier nonlinearities in the wave domain~\cite{11365951}, facilitating integrated sensing and communication (ISAC)~\cite{11157883}, and improving physical-layer security through spatial control~\cite{11050929}.
To optimize the massive number of SIM phase shifts for maximum performance, machine learning (ML) techniques such as deep reinforcement learning (DRL)~\cite{10949617} have been introduced to address the associated non-trivial optimization challenges. \par

However, fully leveraging the deep architectural similarity between SIMs and artificial neural networks (ANNs) to establish a systematic, end-to-end trainable wave-domain processor remains an open challenge.
Each SIM layer and meta-atom structurally mirror an ANN hidden layer and neuron~\cite{11014597}, and phase-shift optimization resembles ANN training. Motivated by this, this article introduces a learnable SIM architecture~\cite{11014597} as a novel ML-in-hardware paradigm for 6G-and-beyond systems. The main contributions are
\begin{itemize}
    \item We propose a learnable SIM architecture to achieve ML-in-hardware for 6G-and-beyond communication systems, reducing digital processing overhead by shifting forward computation from the conventional digital domain to the physical propagation domain.
    The architecture embodies three core principles: training-as-reconfiguration, where architecture training directly configures the SIM; physics-as-structure, which embeds wave propagation constraints into the hardware topology; and hardware-as-algorithm, which leverages the SIM as a physical computing engine.
    \item We consider two SIM-enabled signal processing schemes in MU-MISO uplink systems including multi-user signal separation and jamming signal separation. Through carefully designed SIM phase shifts, the proposed schemes achieve high-performance subchannel orthogonality and reliable communication services. 
    \item We outline pivotal future opportunities for learnable SIM architectures in three key 6G domains: cell-free networks demanding distributed coordination, massive Internet of Things (IoT) systems requiring scalable intelligence, and semantic communications leveraging wave-based processing.
    These interconnected directions collectively chart a course toward ultra-efficient and intelligent wireless infrastructures.
\end{itemize} \par

The remainder of this article is organized as follows. Section~\ref{Sec:II} elucidates the structural analog between SIMs and ANNs, and introduces the proposed learnable SIM architecture. Sections~\ref{Sec:III}-\ref{Sec:IV} design SIM-enabled multi-user separation and jamming signal separation schemes in MU-MISO uplink systems, respectively. Section~\ref{Sec:V} presents pivotal future directions for learnable SIM architectures.
Section~\ref{Sec:VI} summarizes the article. \par

\section{Proposed Learnable SIM Architecture}\label{Sec:II}
This section first overviews the working principle of SIMs and elucidates their structural analogy to ANNs, then introduces the learnable SIM architecture built upon this foundation. \par

\subsection{Fundamentals of Stacked Intelligent Metasurfaces}\label{Sec:II.A}
As shown in Fig.~\ref{fig:II.A.1}(a), a SIM comprises stacked transmissive programmable metasurfaces. Each layer contains a uniform planar array (UPA) of low-cost, passive and tunable meta-atoms, centrally controlled by a smart controller~\cite{11203988}.
When incident signals pass through each layer, each meta-atom acts as a Huygens source~\cite{10515204}, scattering the incident field with a programmable phase shift. The collective field at the next layer results from the coherent superposition of all scattered wavelets, effectively forming a multi-aperture diffraction system whose output is governed by the programmed meta-atom responses and the inter-layer propagation coefficients. \par

By integrating SIMs with base stations (BSs), precoding and combining are executed directly in the EM wave domain, bypassing key bottlenecks in conventional digital transceiver architecture, including
\begin{itemize}
    \item \textbf{Computationally demanding baseband operations}: Traditional MIMO signal processing relies on digital or hybrid precoding/combining, requiring intensive operations like matrix inversion and decomposition~\cite{10515204}.
    \item \textbf{Complex hardware architecture}: Conventional baseband processing handles mixed signals with a large dynamic range, necessitating high-precision, power-hungry DACs/ADCs. Moreover, massive antenna arrays require a prohibitively large number of RF chains. Conversely, the SIM-based transceiver reduces the required RF chains to the number of data streams and relaxes DAC/ADC precision requirement.
\end{itemize} \par

By leveraging EM wave propagation for light-speed parallel processing of all data streams, SIMs enable multi-user signal processing with greatly simplified hardware, superior EE, and reduced computational overhead. These advantages position SIMs as a strong candidate for meeting the demands of massive connectivity and ultra-reliable low-latency communication (uRLLC) in future energy- and cost-efficient wireless systems. \par

\subsection{Construction of SIM-Based Learnable Architecture}\label{Sec:II.B}
\begin{table*}[t]
    \centering
    \caption{\centering{Structural Analogy Between SIMs and ANNs}}
    \small
    \begin{adjustbox}{width=0.98\textwidth}
    \begin{tabular}{|m{3cm}<{\centering}|m{3cm}<{\centering}|m{3cm}<{\centering}|m{6cm}<{\centering}|}
    \hline
        \textbf{Comparison Aspect} & \textbf{SIM} & \textbf{ANN} & \textbf{Contrast with Traditional Baseband Processing}\\ \hline 
        \textbf{Operational Paradigm} & Metasurface layers & Hidden layers & Shift digital computation to wave-domain processing \\ \hline
        \textbf{Basic Unit} & Meta-atoms & Neurons & Passive scatters replace power-hungry components \\ \hline
        \textbf{Fixed Weights} & Inter-layer channels & Fixed parameters & The fixed ``weights'' are governed by EM diffraction laws and channel conditions, remaining constant after fabrication \\ \hline
        \textbf{Trainable weights} & Phase shifts & Adjustable parameters & The SIM phase shifts are trainable parameters, optimized via gradient descent on a nonlinear loss \\ \hline
        \textbf{Optimization Target} & Phase shifts optimization & Network training & The learning process directly translates into the physical reconfiguration of the SIM, rather than the update of abstract software parameters \\ \hline
    \end{tabular}
    \end{adjustbox}
    \label{Table:II.1}
\end{table*}
\subsubsection{Principle of ANNs}\label{Sec:II.B.1}
ANNs are computational models that mimic biological information processing through mathematical representations~\cite{2006Pattern}, capable of learning from data to solve ML tasks such as classification and regression.
A typical ANN architecture comprises an input layer, one or more hidden layers, and an output layer, with information flowing forward through the network.
Each layer contains multiple neurons, which are interconnected across adjacent layers via weight parameters. The output of each neuron is determined by a weighted linear combination of its inputs, followed by a continuous activation function. Network weights are iteratively updated via error backpropagation based on the gradients of a loss function. \par

\subsubsection{Architectural Analogy Between SIMs and ANNs}\label{Sec:II.B.2}
The SIM provides a natural platform for ANN-inspired learning, a framework that transcends conventional optimization by enabling end-to-end and gradient-based learning, as shown in Fig.~\ref{fig:II.A.1}(a). It treats SIM phase shifts as trainable weights and employs backpropagation on a nonlinear loss function, enabling data-driven learning of optimal wavefront transformations. The structural analogies are summarized in Table~\ref{Table:II.1}.
\begin{itemize}
    \item \textbf{From SIM metasurface layers to ANN hidden layers:} Each SIM layer is analogous to an ANN hidden layer, enabling sequential feature extraction in the EM domain~\cite{11014597}.
    \item \textbf{From SIM meta-atoms to ANN neurons:} Each SIM meta-atom serves as a fundamental, tunable unit, analogous to the basic processing elements in the ANN.
    \item \textbf{From SIM channel matrices to ANN fixed weights:} The SIM inter-layer wireless channels, governed by Rayleigh-Sommerfeld diffraction~\cite{10279173}, act as fixed and innate weights. They define the linear combination of wavefronts from all preceding-layer meta-atoms at each current-layer meta-atom.
    \item \textbf{From SIM phase shifts to ANN trainable weights:} Programmable meta-atom responses serve as trainable parameters, performing complex-valued linear rotations on the incoming signals. Their gradient-based optimization driven by a nonlinear loss function, i.e., the squared Euclidean distance between the normalized symbols, enables the SIM to learn complex wavefront transformation for target objectives.
    \item \textbf{From SIM phase optimization to ANN network training:} Gradient-based optimization of phase shifts via error backpropagation mirrors ANN training, with updates physically realized as meta-atom reconfiguration to minimize a task-specific loss function, thereby enhancing performance metrics such as average signal-to-interference-pulse-noise ratio (SINR) and sum rate.
\end{itemize} 
These structural similarities establish the SIM as a physical ML platform, enabling light-speed forward propagation and error backpropagation training of its hardware parameters.
The end-to-end mapping remains linear, and the benefit of the multilayer structures lies in constrained linear transformation rather than universal nonlinear approximation. \par

\subsubsection{Learnable SIM-based ML Paradigm}\label{Sec:II.B.3}
The proposed learnable SIM architecture can embody a novel \textit{ML-in-hardware} concept, effectively reducing the digital processing load by executing forward propagation in the physical domain of EM waves.
Its core principles are summarized as follows
\begin{itemize}
    \item \textbf{Training-as-reconfiguration}: The error backpropagation algorithm directly optimizes the physically tunable SIM phase shifts, translating model updates into hardware reconfiguration.
    \item \textbf{Physics-as-structure}: Rather than being arbitrarily designed, the structure is intrinsically defined by the EM propagation laws and the spatial arrangement of SIM layers.
    \item \textbf{Hardware-as-algorithm}: The SIM is a physical entity whose inherent wave propagation process constitutes forward propagation, merging algorithmic function with hardware implementation.
\end{itemize} \par

Consequently, the learnable SIM architecture paves the way for hardware-efficient and scalable intelligent wireless communication systems. \par

\subsection{Training of Learnable SIM Architecture}\label{Sec:II.C}
The transmission frame is structured into three primary phases, called pilot transmission, coefficient configuration, and data transmission~\cite{9744412}. 
The $U$ pilot symbols serve the dual role of providing the necessary channel state information (CSI) of SIM-user channels and simultaneously forming the dataset for training the SIM parameters, without introducing additional estimation overhead.
During the coefficient configuration phase, the SIM phase shifts are updated by backpropagation with $U$ pilot symbols to minimize the loss function, which mirrors ML's training stage. Once trained, the SIM executes the learned transformation at light speed during data transmission. \par

\subsubsection{Loss Function}
The loss function is defined as the squared Euclidean distance between the normalized received and transmitted symbol vectors with each scaled by its own Euclidean norm ($\ell_2$-norm). This normalization projects the signals onto a unit hypersphere, mitigating large-scale fading effects, making it particularly effective for phase-shift-keying (PSK) modulation schemes, e.g., quadrature PSK (QPSK), and inherently extensible to higher-order PSK. Illustrations for different signal processing scenarios are provided in Figs.~\ref{fig:II.A.1}(b)-(c). \par

\subsubsection{SIM Phase Shifts Training Process}
During the coefficient configuration phase, the SIM phase shifts are trained over $T$ episodes via gradient-based error backpropagation to minimize the loss function. \par

The algorithm is initialized with the following inputs: the SIM-user channels $\mathbf{H}$, the inter-layer SIM channels $\{\mathbf{W}_l\}_{l=2}^{L}$, a set of $U$ pilot symbols $\{\mathbf{s}_{\mu}\}_{\mu = 1}^{U}$, initial learning rate and decay rate $\eta_{0}$, $\beta$, SIM layers $L$, number of episodes $T$, tolerant error $\epsilon$, initial SIM phase shifts $\{\mathbf{\Phi}_{l}\}_{l=1}^L$. Each episode $t$ then executes the following procedure
\begin{itemize}
    \item \textbf{Update learning rate}: the learning rate $\eta$ follows an adaptive strategy, where it undergoes an exponential decay $\eta^{t} = \beta^{t-1}\cdot\eta_{0}$ over training episodes, starting with $\eta_{0}$ in the first episode.
    \item \textbf{Update SIM phase shift layer-by-layer}: SIM phase shifts are updated layer-by-layer according to the error backpropagation algorithm
        \begin{enumerate}
            \item For each SIM layer ($l = 1, \cdots, L$)
            \item Compute the mini-batch gradient: For each pilot slot in the batch, the gradient of the loss function is computed with respect to the phase shift of every meta-atom within the current layer. These per-slot gradients are then accumulated across the entire batch.
            \item Update parameters: All phase shifts in the current SIM layer are simultaneously updated using the average mini-batch gradient and the current learning rate $\eta^{t}$.
        \end{enumerate}
    \item \textbf{Determine the termination conditions}: The algorithm terminates when it reaches the maximum episode $T$ or the difference in loss function between adjacent episodes is less than the tolerant error $\epsilon$.
\end{itemize} \par

Conventional non-convex optimization methods such as alternating optimization typically rely on high-complexity matrix operations, e.g., singular value decomposition, and may suffer from slow convergence or local optima. Conversely, our data-driven approach directly minimizes the per-slot squared error, an end-to-end metric aligned with symbol recovery, that scales gracefully with user number. Meanwhile, systematic integration of mini-batch gradient descent and adaptive learning rate jointly provides more stable and efficient convergence. \par

The aforementioned CSI acquisition and training process are conducted in the digital domain to compute the optimal phase shifts. Once configured, the SIM executes the forward propagation entirely within the analog EM domain at the speed of light. This training-as-reconfiguration and hardware-as-algorithm paradigm combines the advantages of tractable digital optimization with the ultra-efficiency of wave-based analog computing. \par

\section{SIM-Empowered Subchannel Orthogonality}\label{Sec:III}
This section leverages the learnable SIM architecture to design a novel scheme for multi-user separation in MU-MISO uplink systems, and its effectiveness and superior performance are demonstrated through corresponding simulations. \par

\subsection{Use Case Scenario of MU-MISO Communications}\label{Sec:III.A}
\subsubsection{SIM-Enabled Multi-User Signal Separation Method}\label{Sec:III.A.1}
This section proposes a SIM-enabled MU-MISO system, where the SIM is integrated with the BS as the combiner to effectively distinguish multi-user uplink signals, as illustrated in Fig.~\ref{fig:II.A.1}(b).
The proposed learnable SIM architecture comprises $L$ layers, and each layer contains $N = N_{x}N_{y}$ meta-atoms, arranged in a UPA configuration~\cite{11157883}.
The BS deploys an $M$-element uniform linear array (ULA) to receive signals from $K$ single-antenna users.
A subset of $K$ antennas is selected from the $M$ available to create a dedicated one-to-one mapping with users, facilitating efficient quasi-orthogonal subchannels~\cite{10949617}.
By treating the SIM as a physical computational graph, we perform a regression task on the received multi-user mixed signals. This process inherently assigns each user to a dedicated antenna, enabling the reconstruction of the original data symbols transmitted by each user at the BS for efficient end-to-end multi-user communication. \par

\begin{figure*}[!t]
	\centering{}\includegraphics[width=7.2in]{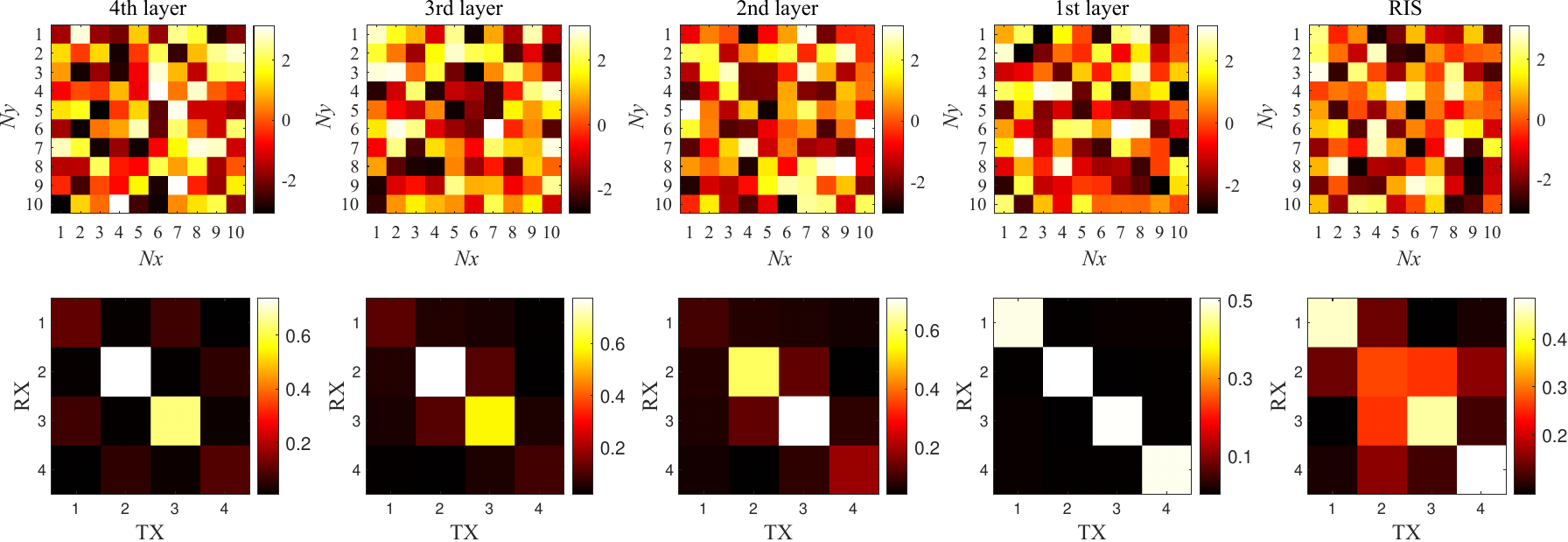}
	\caption{Visualization of end-to-end quasi-orthogonal subchannel formation via SIM-based wavefront processing, with $K = 4$ users, and $N = 100$ meta-atoms, compared with conventional RISs.}
	\label{Sec:III.B.1.1}
\end{figure*}
\subsubsection{The learnable SIM Parameter Optimization}\label{Sec:III.A.2}
During the coefficient configuration phase, the SIM phase shifts are updated layer-by-layer via gradient-based error backpropagation over $T$ episodes, using $U$ pilot symbols previously exchanged between the BS and users during the preceding pilot transmission period. 
The loss function is defined as the squared Euclidean distance between the normalized received and transmitter symbols.
The corresponding optimization problem is formulated to minimize this loss function, as shown in Fig.~\ref{fig:II.A.1}(b), thereby striving to approximate the directional fidelity of the original transmitted signal vectors as closely as possible. \par

Additionally, during the optimization of $N$ meta-atoms in the $l$-th SIM layer, a mini-batch gradient descent approach is employed over $U$ time slots, rather than a slot-by-slot update, to balance convergence accuracy and computational complexity, substantially reducing computational complexity from $\mathcal{O}(N^{2}MU)$ to $\mathcal{O}(NMU)$. Concurrently, an adaptive learning rate with a decay strategy is adopted to dynamically balance convergence speed, precision, and training stability. \par

This SIM-enabled ML paradigm avoids computationally intensive operations, including matrix multiplication, decompositions, inversion, and complex hardware components such as numerous RF chains and high-precision ADC converters, which are essential in traditional baseband signal combining~\cite{10515204}. By executing light-speed signal operations through EM wave propagation, it achieves unprecedented hardware efficiency and processing speed, harnessing the innate parallelism of wave physics~\cite{10534211}. \par

Notably, although demonstrated here for uplink signal separation, the proposed paradigm can also be configured as a downlink transmit precoder to manage MUI directly in the wave domain, showcasing its versatility as a foundational 6G-and-beyond transceiver technology. \par

\subsection{The Role of SIM in Subchannel Orthogonality}\label{Sec:III.B}
This subsection presents a series of simulations to evaluate the impact of SIM's key architecture parameters, such as the number of layers and meta-atoms per layer, on the performance of multi-user uplink signal separation. \par

\subsubsection{Simulation Setup}
\begin{itemize}
    \item \textbf{Channel models}: The BS-SIM and inter-layer SIM channels are modeled via Rayleigh-Sommerfeld diffraction theory~\cite{10279173}, while the SIM-user channels follow a correlated three-dimensional Rician fading model~\cite{10949617}, which are shown in Fig.~\ref{fig:II.A.1}(b).
    \item \textbf{SIM parameters}: The SIM comprises $L$ layers, each with $N_x \times N_y$ half-wavelength spaced meta-atoms, i.e., $\Delta = \lambda/2$. Its total thickness is $T_{\mathrm{SIM}} = 10\lambda$, yielding an inter-layer spacing of $D_L = T_{\mathrm{SIM}} / (L-1)$.
\end{itemize}

\subsubsection{Impact of SIM Key Parameters}\label{Sec:III.B.1}
Fig.~\ref{Sec:III.B.1.1} visually illustrates the evolution of the wavefront and phase configuration through successive SIM layers in an uplink MU-MISO system, compared to conventional RISs. As the wavefront propagates through stacked metasurfaces, the diagonal elements of the equivalent channel matrix are progressively enhanced, while the off-diagonal elements are effectively suppressed. Specifically, after the final layer, the normalized equivalent channel exhibits a 25.77\% increase and a 78.11\% reduction in average diagonal and non-diagonal element strength with that after only the first layer, respectively.
This trend indicates the gradual establishment of independent subchannels for each user. Compared to a single-layer RIS~\cite{cao2024intelligent}, the improvements are 20.26\% and 94.77\%, respectively, with diagonal element variance reduced by 24.89 dB. This transformation stems from substantially increased DoFs, which are afforded by the multilayer SIM architecture and enable the projection of the MUI onto an orthogonal subspace.
Consequently, efficient end-to-end quasi-orthogonal subchannels are established between each user and its designated antenna, yielding effective MUI suppression and substantial system performance gain. \par

The convergence of the learnable SIM architecture under different hyperparameters is evaluated to quantify the contribution of the proposed learning mechanism. Figs.~\ref{Sec:III.B.2.1}(a)-(b) plot the average loss, with $\pm 1$ standard deviation bands, vs. episodes for initial learning rate $\eta_0 \in \{0.75, 0.85, 0.95\}$ with $\beta = 0.985$, and for $\beta \in \{0.97, 0.98, 0.99\}$ with $\eta_0 = 0.8$, respectively, given $K = 4$, $L = 5$, and $N = 64$. The results demonstrate a steady decrease and consistent convergence in all settings. The terminal relative loss changes are 0.37\%, 0.45\%, 0.41\% for $\eta_0 = 0.75, 0.85, 0.95$, and 0.03\%, 0.13\%, 0.78\% for $\beta = 0.97, 0.98, 0.99$, respectively.
Higher $\eta_0$ or $\beta$ yields a faster initial loss drop, yet introduces larger oscillations, which are ultimately dampened, leading to significantly lower final loss. Specifically, $\eta_0 = 0.95$ achieves a final loss of 61.86\% lower than $\eta_0 = 0.75$, and $\beta = 0.99$ reduces the loss by two orders of magnitude versus $\beta = 0.97$.
This confirms that the training mechanism itself is effective, robust, and controllable, leading to reliable convergence at a steadily low loss. \par

Fig.~\ref{Sec:III.B.2.1}(c) depicts constellation diagrams as the number of meta-atoms per SIM layer increases from $N = 25$ to 64, under the configuration $K=4$ users, and $L=5$ layers.
The received symbols progressively converge to ideal QPSK positions.
Quantitatively, as $N$ grows to 36, 49, and 64, the mean squared errors (MSEs) relative to the ideal QPSK symbols reduce to $1.092\times10^{-4}$, $2.147\times10^{-6}$, and $1.674 \times 10^{-8}$, respectively, and decrease approximately two, four, and six orders of magnitude compared to the MSE of $1.456\times10^{-2}$ at $N=25$.
This improvement stems from the enhanced representational capacity enabled by the wider SIM architecture, which leads to stronger multi-user separation capability. \par

\subsubsection{Effect of SIM on MUI Suppression}\label{Sec:III.B.2}
\begin{figure}[!t]
	\centering{}\includegraphics[width=3.5in]{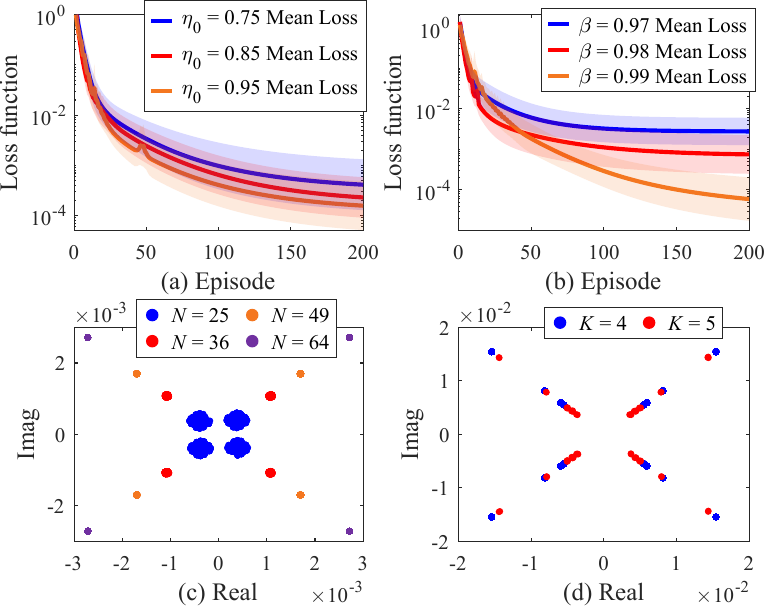}
	\caption{Average loss functions vs. episodes for different $\eta_0$ and $\beta$; constellation diagrams for different values of $N$, and invariant angles and ranged distance.}
	\label{Sec:III.B.2.1}
\end{figure}
Fig.~\ref{Sec:III.B.2.1}(d) shows the constellation diagrams during the data transmission period for $K = 4$ and $K = 5$, where user distances to the BS are randomly scaled by factors between 0.5 and 1.5 while their elevation and azimuth angles remain invariant.
Due to effective MUI suppression achieved during the coefficient configuration phase, the trained SIM successfully projects interference signals into an orthogonal subspace. Under noise-free conditions and invariant user angles, the received constellation points exhibit phase angles that match the ideal QPSK references.
Furthermore, as the transmission distance varies, the constellation points distribute linearly along the phase directions rather than clustering at singular locations.
This phenomenon confirms that the proposed approach enables the SIM to adaptively project MUI into an orthogonal subspace based on relative user positions, thus achieving effective MUI management and demonstrating robust spatial interference suppression capabilities. \par

\section{SIM-Empowered Jamming Separation}\label{Sec:IV}
This section explores a further application of the learnable SIM architecture, i.e., the separation of communication and jamming signals, in MU-MISO uplink systems. Corresponding simulations validate the effectiveness of the proposed scheme in classifying and distinguishing between the desired communication and jamming signals. \par

\subsection{Use Case Scenario of Wireless Jamming}\label{Sec:IV.A}
Based on the SIM-enabled MU-MISO system proposed in Section~\ref{Sec:III}, this section further introduces an attacker, called Mallory, to degrade the communication performance of $K$ legitimate users by transmitting jamming signals during the transmission frame~\cite{11203988}, as depicted in Fig.~\ref{fig:II.A.1}(c). We examine the role of the learnable SIM architecture in this adversarial scenario, exploring mitigation strategies to counteract jamming signals. Our aim is to reveal the inherent advantages of the SIM architecture in enhancing the robustness of the system against jamming attacks. \par

\subsubsection{Jamming Attack on MU-MISO System}
Mallory can continuously transmit additive white Gaussian noise (AWGN) jamming signals or even adversarial perturbations toward the BS to disrupt the correct demodulation of multi-user communication signals~\cite{11203988}, thereby degrading overall system performance, such as the average SINR and sum rate. This real-time jamming attack does not require CSI of legitimate users, which not only enhances its effectiveness and execution feasibility but also dramatically increases the difficulty of conventional jamming separation techniques. \par

\subsubsection{SIM-Enabled Anti-Jamming Architecture}
The proposed SIM-enabled anti-jamming architecture addresses this challenge by training SIM parameters during the coefficient configuration period using $U$ received pilot signals that include jamming transmissions.
The trained architecture learns to configure the SIM for spatial filtering, effectively suppressing both jamming signals and MUI by forming deep nulls in their respective spatial directions, while preserving the desired user signals.
The squared Euclidean distance between the normalized received and transmitted symbols is also adopted as the loss function, as illustrated in Fig.~\ref{fig:II.A.1}(c), and the learnable SIM architecture is also optimized using a mini-batch gradient-based error backpropagation algorithm with adaptive learning hyperparameters. \par

\subsection{The Role of SIM in Jamming Separation}\label{Sec:IV.B}
\begin{figure}[!t]
	\centering{}\includegraphics[width=3.5in]{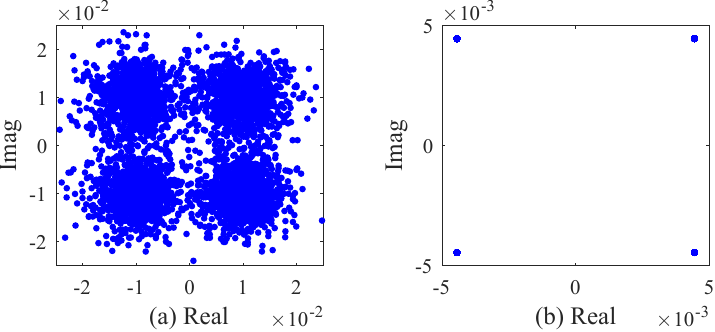}
	\caption{Constellation diagrams with jamming-agnostic and jamming-aware SIM configurations during the training period.}
	\label{fig:IV.B.1}
\end{figure}

This subsection presents simulations of the improved SIM architecture for the separation of communication and jamming signals under jamming attacks. The adversary Mallory is modeled as a reactive jammer randomly located among users. Its jamming signal is generated as AWGN with variance 0.5, which is then amplified by a transmitter with random time-varying power. The simulations demonstrate the effectiveness of the proposed architecture in preserving reliable signal demodulation at the BS and maintaining system performance matrices, such as the symbol error rate (SER) and sum rate. \par

\subsubsection{Monte Carlo Settings and Robustness Tests}
We employ a Monte Carlo method over two levels of randomness. First, $2^5$ independent SIM-user channel realizations $\mathbf{H}=[\mathbf{h}_1, \mathbf{h}_2, \cdots, \mathbf{h}_K] \in \mathbb{C}^{N \times K}$ are generated.
For each, the SIM phase shifts are optimized. Then, over $2^{15}$ time slots per signal-to-noise ratio (SNR) point, data symbols are transmitted with AWGN added. The final performance curves are the average across all channel realizations, ensuring statistical robustness. \par

\subsubsection{Simulation Results}
Fig.~\ref{fig:IV.B.1} presents noise-free constellation diagrams comparing jamming-agnostic and jamming-aware SIM configurations under a jamming attack from Mallory.
In the jamming-agnostic case, i.e., Fig.~\ref{fig:IV.B.1}(a), the receiver fails to effectively separate jamming from multi-user communication signals, resulting in received symbols that exhibit circularly symmetric complex Gaussian distributions around each ideal QPSK point, with an MSE of $17.741\times10^{-2}$. Conversely, when the proposed jamming-aware SIM is employed, the constellation points concentrate closely around the ideal QPSK references, as shown in Fig.~\ref{fig:IV.B.1}(b). This demonstrates that during coefficient configuration, by minimizing the squared Euclidean distance between normalized received and transmitted symbols, the optimized SIM aligns received symbol phases with original transmission phases while projecting jamming signals into an orthogonal subspace, thereby enabling correct demodulation. \par

\begin{figure}[!t]
	\centering{}\includegraphics[width=3.5in]{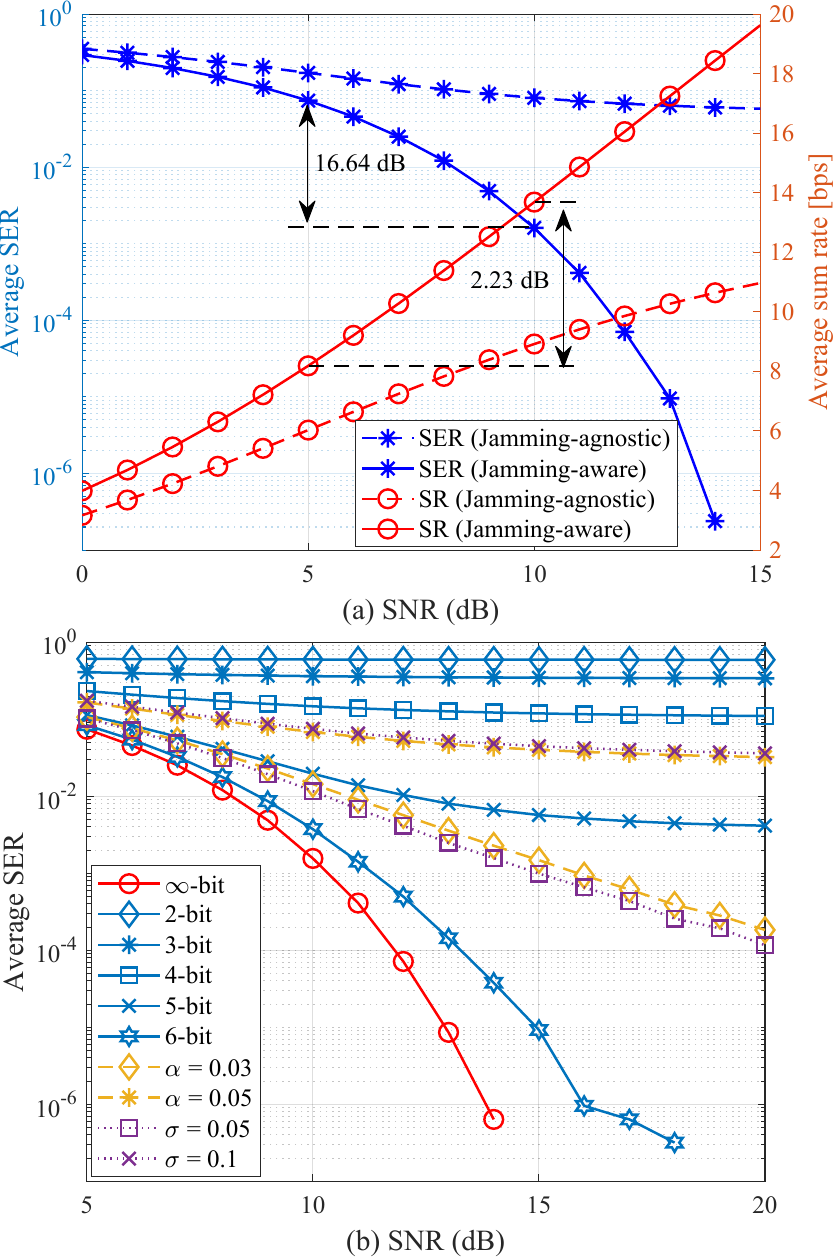}
	\caption{(a) SER and SR comparison with jamming-agnostic and jamming-aware SIM configurations across various SNR, with $K = 4, N = 100, L=6$; (b) SER performance under hardware imperfections.}
	\label{fig:IV.B.2}
\end{figure}
Fig.~\ref{fig:IV.B.2}(a) quantitatively compares the SER and sum rate performance between jamming-agnostic and jamming-aware SIM configurations under varying SNRs. When the jamming-aware SIM is employed, as SNR increases from 5 dB to 10 dB, SER decreases by 16.64 dB and sum rate improves by 2.23 dB.
Conversely, with the jamming-agnostic SIM, SER convergences to a high floor above $5.8 \times 10^{-2}$, and sum rate grows sluggishly with SNR.
This is because jamming severely corrupts the channels between each user and its corresponding BS antenna, preventing the formation of an end-to-end quasi-orthogonal subchannel, which leads to a received signal distribution approximating circularly symmetric complex Gaussian and makes correct demodulation difficult. \par

Furthermore, hardware discreteness is modeled by quantizing each SIM phase shift $\varphi_{l,n}$ to its nearest $B$-bit discrete level $\tilde{\varphi}_{l,n}$, i.e., $\tilde{\varphi}_{l,n} = \left\lbrack \frac{\varphi_{l,n}}{2\pi/2^{B}} \right\rbrack \times \frac{2\pi}{2^{B}}$, where $[\cdot]$ denotes the round operation. As illustrated in Fig.~\ref{fig:IV.B.2}(b), low resolution (2-4-bit) induces severe quantization noise, resulting in nearly flat SER curves. At 5-bit, SER decreases with SNR but floors around $4 \times 10^{-3}$. Performance with 6-bit resolution approximately approaches that of continuous phase shifts, achieving SER below $10^{-6}$ at 16 dB SNR, indicating a practical engineering threshold to overcome quantization effects. \par

Real-world metasurfaces also exhibit inter-element coupling and non-ideal phase responses. The former is captured by a banded symmetric Toeplitz coupling matrix $\mathbf{C}$~\cite{7807222}, where the interaction between elements decays exponentially with index distance as $c_{i,j} = \alpha^{|i-j|}$ for $|i-j| \leq 5$ and 0 otherwise, with coupling coefficient $\alpha \in \{0.03, 0.05\}$. The latter is modeled as von Mises-distributed noise~\cite{9417454} with mean 0, standard deviation $\sigma \in \{0.05, 0.1\}$, range $[- \pi, \pi)$, added to each optimized phase shift.
As shown in Fig.~\ref{fig:IV.B.2}(b), mild distortion levels ($\sigma = 0.05$ or $\alpha = 0.03$) allow near-ideal SER decay, whereas higher distortion levels ($\sigma = 0.1$ or $\alpha = 0.05$) induce a distinct error floor above $3.6 \times 10^{-2}$ and $3.2 \times 10^{-2}$, respectively, quantifying the performance penalty.
This underscores the necessity of advancing from idealized models toward measurement-calibrated joint EM-communication frameworks in future research~\cite{11157883}. \par

\section{Open Issues and Future Directions}\label{Sec:V}
This section discusses future research directions for learnable SIM architectures in three key application domains: cell-free communications, massive IoT, and semantic communications. \par

\subsection{SIM-Enabled Cell-Free Communications}\label{Sec:V.A}
Leveraging superior capabilities in interference management, SIMs are poised to significantly enhance cell-free systems in coverage and EE. However, their practical realization requires addressing key challenges in distributed training across access points (APs) and maintaining robustness in dynamic environments. 
A core open problem is designing communication-efficient distributed training protocols that function with highly local CSI.
Future work should focus on scalable coordination algorithms to overcome the pilot contamination problem and enable efficient distributed precoding with limited inter-AP CSI exchange. 
Another key challenge is developing SIM parameter topology. Addressing these will be crucial to alleviate the fronthaul burden and latency for truly scalable cell-free networks. \par

\subsection{Collaborative SIM Training for Massive IoT}\label{Sec:V.B}
SIMs present a paradigm-shifting solution for 6G massive IoT, leveraging wave-based analog computing to achieve unprecedented EE in handling massive device connectivity.
Future research should therefore address two pivotal problems. The first is designing asynchronous event-driven SIM training algorithms that efficiently accommodate the sporadic and heterogeneous traffic patterns of massive IoT devices. The second is developing specialized federated intelligence framework for reliable and communication-efficient model synchronization across vast networks of resource-constrained SIM nodes. Solving these problems is key to realizing dynamic adaptability and meeting rigorous reliability requirements for industrial IoT applications. \par

\subsection{Semantic Communication System Based on SIM}\label{Sec:V.C}
SIMs provide an ideal hardware platform for semantic communications~\cite{11050929} by enabling direct wave-domain semantic feature extraction, overcoming the computational overhead of digital approaches.
This achieves task-oriented semantic transmission with ultra-low power and minimal latency, while also simplifying transceiver architectures and enhancing low-SNR robustness.
The key technical challenges for this integration lie in
developing adaptive training algorithms for dynamic wireless channels, and maintaining semantic accuracy under hardware imperfections and manufacturing tolerances. Future work should focus on joint optimization of SIM configurations and semantic encoders to fully realize this synergy. \par

\section{Conclusion}\label{Sec:VI}
This article presents a novel learnable SIM-based ML paradigm for 6G-and-beyond systems by proposing the learnable SIM architecture. The hardware-efficient architecture is designed to process signals directly in the EM wave domain by leveraging the structural analogy between SIMs and ANNs, thereby implementing key functions, such as multi-user and jamming signal separation. \par

Through two use cases in multi-user communication
scenarios, it shows how the SIM works to achieve complicated signal processing functions.
In the future, we will explore SIM implementation issues across three application domains, including distributed coordination in cell-free networks, scalable intelligence for massive IoT deployments, and efficient wave-domain processing for semantic communications. \par

\bibliographystyle{IEEEtran}
\bibliography{ref.bib}

\begin{IEEEbiographynophoto}{Hetong Wang}
received the B.S. and M.S. degrees from the School of Telecommunications Engineering, Xidian University, Xi'an, China, in 2020 and 2023, respectively. She is currently pursuing the Ph.D. degree at the School of Information and Communication Engineering, Beijing University of Posts and Telecommunications (BUPT), Beijing, China. Her current research interests include physical layer security, reconfigurable intelligent surface, stacked intelligent metasurface, and machine learning. She has served as a reviewer for many high-impact IEEE journals, such as IEEE WCM, IoT, WCL, CL, OJCOMS.
\end{IEEEbiographynophoto}

\begin{IEEEbiographynophoto}{Yashuai Cao}
received the B.E. and Ph.D. degrees in communication engineering from Chongqing University of Posts and Telecommunications (CQUPT) and Beijing University of Posts and Telecommunications (BUPT), China, in 2017 and 2022, respectively. From 2022 to 2023, he was a lecturer in the Department of Electronics and Communication Engineering, North China Electric Power University (NCEPU), Baoding. From 2023 to 2025, he was a Postdoctoral Research Fellow with the Department of Electronic Engineering, Tsinghua University, Beijing, China. He is currently a Distinguished Associate Professor with the School of Intelligence Science and Technology, University of Science and Technology Beijing, Beijing, China. His research interests include Stacked Intelligent Metasurface, Environment-Aware Communications, and Channel Knowledge Map.
\end{IEEEbiographynophoto}

\begin{IEEEbiographynophoto}{Tiejun Lv}
received the M.S. and Ph.D. degrees in electronic engineering from the University of Electronic Science and Technology of China (UESTC), Chengdu, China, in 1997 and 2000, respectively. From January 2001 to January 2003, he was a Postdoctoral Fellow at Tsinghua University, Beijing, China. In 2005, he was promoted to Full Professor at the School of Information and Communication Engineering, Beijing University of Posts and Telecommunications (BUPT). From September 2008 to March 2009, he was a Visiting Professor with the Department of Electrical Engineering at Stanford University, Stanford, CA, USA. He is the author of four books, one book chapter, more than 160 published journal papers and 200 conference papers on the physical layer of wireless mobile communications. His current research interests include signal processing, communications theory and networking. He was the recipient of the Program for New Century Excellent Talents in University Award from the Ministry of Education, China, in 2006. He received the Nature Science Award from the Ministry of Education of China for the hierarchical cooperative communication theory and technologies in 2015 and the Shaanxi Higher Education Institutions Outstanding Scientific Research Achievement Award in 2025.
\end{IEEEbiographynophoto}
\end{document}